\documentclass[physrev,graphicx,twocolumn,superscriptaddress,reprint]{revtex4-2}
\usepackage{vacuum_arc_plasma_surface}

\usepackage[pdftex]{hyperref}

\usepackage{academicons}
\usepackage{orcidlink}

\draft 

\begin{document}


\title{Coupled simulation of plasma-surface interactions during early stages of vacuum arcing} 



\newcommand{\helsinki}{
  Helsinki Institute of Physics and Department of Physics, P.O. Box 43 (Pietari Kalmin Katu 2), FI-00014 University of Helsinki, Finland
}
\newcommand{\tartu}{
  Institute of Technology, University of Tartu, Nooruse 1, 50411 Tartu, Estonia
}
\author{Roni Koitermaa \orcidlink{0000-0001-9814-7358}}
\email[]{roni.koitermaa@helsinki.fi}
\email[]{roni.koitermaa@iki.fi}
\affiliation{\helsinki}
\affiliation{\tartu}

\author{Andreas Kyritsakis \orcidlink{0000-0002-4334-5450}}
\affiliation{\tartu}

\author{Tauno Tiirats \orcidlink{0000-0002-1256-7392}}
\affiliation{\tartu}

\author{Flyura Djurabekova \orcidlink{0000-0002-5828-200X}}
\email[]{flyura.djurabekova@helsinki.fi}
\affiliation{\helsinki}

\author{Veronika Zadin \orcidlink{0000-0003-0590-2583}}
\email[]{veronika.zadin@ut.ee}
\affiliation{\tartu}


\date{\today}

\begin{abstract}
  We describe fully coupled simulations that bridge atomistic cathode dynamics and plasma formation during the earliest stages of vacuum arcing. The model combines molecular dynamics, finite element electrothermal calculations, electron emission and particle-in-cell plasma simulations via dynamic transfer of particles between the surface and plasma domains. Simulations of Cu nanoprotrusions reveal two routes to thermal runaway: direct Joule heating-driven instability and a novel nanoparticle-assisted mechanism, where detached nanoparticles generate neutral vapor that becomes ionized.
\end{abstract}


\maketitle 



%
%

%

\fig{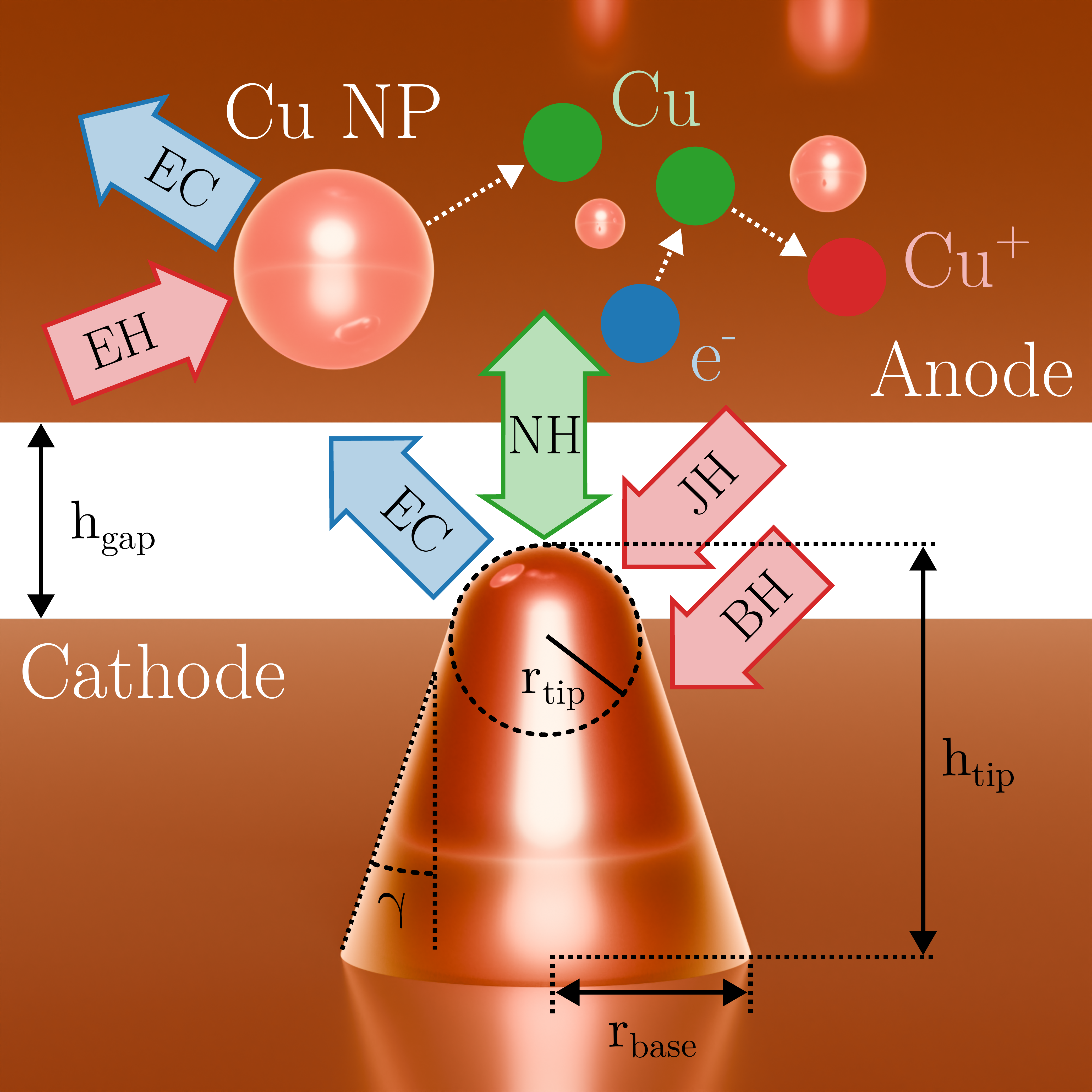}{Schematic of the simulated system and the different processes of vacuum arcing included in the model. Emitted electrons ionize evaporated neutral Cu atoms that originate from the surface or from nanoparticles (NPs), forming Cu$^+$ ions, which bombard the cathode. Heat fluxes include: Joule heat (JH), Nottingham heat (NH), bombardment heat (BH), evaporative cooling (EC) and electron heating (EH).\label{fig:system}}{0.7}

Despite studies spanning decades, vacuum arcs, also known as vacuum breakdowns (VBD), still elude detailed understanding concerning their fundamental mechanisms~\cite{wue25}. Vacuum arcing is a limiting factor in high-field applications such as vacuum interrupters, fusion reactors and particle accelerators, including the proposed Compact Linear Collider (CLIC)~\cite{cli26}, Future Circular Collider (FCC)~\cite{fcc25} and Muon Collider~\cite{muc24}. Current understanding connects the vacuum arcing process to enhancement of the electric field on the surface (field enhancement factor $\beta$), localized heating, creation of neutral vapor and ionization of this vapor to form a conducting plasma. Initial field enhancement has been proposed to originate from the formation of nanoprotrusions on electrode surfaces, which would increase the surface $\beta$, allowing electron emission to start~\cite{jen08, nor12, poh13, zad14, jan20, eng19}. Recently, significant modeling effort has gone towards understanding these processes and the connections between them, namely using plasma simulations~\cite{tim15} and multiscale/multiphysics simulations of emitting nanoprotrusions under field~\cite{dju11, par11, ves16, ves18, kyr18, ves20}. These processes are illustrated in Fig.~\ref{fig:system}, which shows a Cu cathode surface with a nanoprotrusion, separated by a vacuum gap from the anode. Once ionization of neutral vapor above the cathode starts, ions would be accelerated towards the surface, creating a feedback loop that contributes towards the creation of additional vapor due to sputtering and heating. Additionally, the charge distribution resulting from ions can enhance the surface electric field. Cathode surface and plasma processes are significantly connected in vacuum arcing, but so far no model has fully coupled them together in a unified framework.

Previous simulations have shown that nanoparticles (NPs) can be detached from nanoprotrusions under field~\cite{ins04, kyr18, ves20}. The ejection of such fragments could serve as contributors to the formation of plasma~\cite{ins10}. The creation of macroparticles during vacuum arcing has been experimentally observed, likely originating from the cathode spot due to plasma-surface interactions~\cite{box92, and99}. It has also been proposed that the presence of microparticles could lead to a distinct breakdown mechanism via cascade discharge~\cite{liu25}.

A simulation model is developed that couples both the processes occurring in the cathode, as well as in the plasma above the cathode surface. The goal for such a model is to better capture plasma-surface interactions that occur during vacuum arc initiation. Atomistic simulation of the cathode is performed using molecular dynamics (MD)~\cite{kyr18, ves20}, while the plasma is simulated using the particle-in-cell method (PIC)~\cite{tim15, koi24} following previous work. The current simulation model couples these two systems together using a process of particle transfer.

Simulation of the cathode is performed using the FEMOCS framework~\cite{ves20}. This is a multi-scale, multi-physics model that simulates metal surfaces under high electric fields, combining electron emission, electric current, heat and atomistic calculations. The Poisson's, continuity and heat equations are solved using the finite element method (FEM), implemented via the deal.II library~\cite{dea21}. The electric field in the vacuum between the cathode and anode is solved from Poisson's equation
\be
\nabla^2 \phi = -\frac{\rho}{\eps_0},
\ee
based on Dirichlet or Neumann boundary conditions, where $\phi$ is electric potential, $\rho$ is the charge density distribution in the vacuum resulting from electrons and ions and $\eps_0$ is the vacuum permittivity. Electron emission ranging from cold field emission to thermionic emission is calculated using the GETELEC library~\cite{kyr17}. Current and temperature distributions are solved in the cathode bulk. Current is solved from the continuity equation
\be
\nabla \cdot (\sigma(T) \nabla \phi) = 0,
\ee
where $\sigma(T)$ is electrical conductivity and current density is given by $\vb J = \nabla \phi$. The bottom of the cathode is grounded and current density on the surface of the cathode is given by the electron emission current $\vb J_e$. The temperature distribution $T$ is solved from the heat equation
\be
C_V \frac{\partial T}{\partial t} = \nabla \cdot (\kappa(T) \nabla T) + P_J,
\ee
where $C_V$ is the volumetric heat capacity and $\kappa(T)$ is the thermal conductivity~\cite{nat74, sch01, yar09}. The main heat sources for the heat equation include Joule heating $P_J$ due to current and Nottingham heating $P_N$ of the surface due to emission. Meshes for FEM are generated automatically from atom positions using the TetGen library~\cite{han15}. Molecular dynamics (MD) simulated using LAMMPS~\cite{lam22} is used to advance the positions of atoms, which results in deformation of the cathode surface, influencing the surrounding electric field, emission and heating. Surface charges are distributed to surface atoms based on Gauss's law, which results in forces due to the external electric field~\cite{ves20}.

Plasma simulation is performed in FEMOCS using the particle-in-cell method (PIC)~\cite{tsk07}, as described in previous work~\cite{tim15,koi24}. Superparticles (SPs), are injected into the PIC system, such as electrons resulting from emission and neutrals from evaporation or sputtering. These particles influence the electric field and move as a result of electric forces. Particles undergo collisions, implemented as Monte Carlo collisions (MCC)~\cite{tak77}, such as elastic, Coulomb, ionization, charge exchange and recombination. The collision probability $P$ is~\cite{vah95}
\be
P = 1 - \exp(-u n \sigma(E) \Delta t_\x{PIC}),
\ee
where $u$ is the relative velocity of colliding particles, $n$ is the number density, $\sigma(E)$ is the cross section~\cite{tra77, pin85, gre86, bol94, fre90} and $\Delta t_\x{PIC}$ is the PIC time step.
This can result in the formation of ions at various charge levels, which get accelerated towards the cathode surface and bombard it, causing heating or sputtering. A more detailed description of the plasma interactions is given in previous work~\cite{koi24}. A detailed description of the simulation model is given in Appendix~\ref{sec:model}.

The simulated system (shown in Fig.~\ref{fig:system}) consists of Cu plate electrodes with a gap of $h_\x{gap} = 1200 a$ (where $a = 3.64 \us{Å}$ is the Cu lattice parameter). The width of the simulation box is $w = 800 a$ and the depth of the cathode bulk is $h_\x{bulk} = 200 a$. On the cathode surface, a conical nanoprotrusion is assumed to be present with a height $h_\x{tip} = 100 \us{nm}$, top radius $r_\x{tip} = 1 \us{nm}$ and bottom radius $r_\x{base} = r_\x{tip} + (h_\x{tip} - r_\x{tip}) \tan(\gamma)$, with opening angle $\gamma \approx 1.7 \degs$. The bottom 100~Å is flared to three times the bottom radius. The entire nanoprotrusion is simulated atomistically, while the flat cathode bulk is assumed to be static. The nanotip consists of 346764 Cu atoms, with the FCC lattice oriented such that the cathode top surface is \{100\}.

Simulations are performed at various external fields of $F_\x{ext} = 450 \us{MV}/\un{m}$ ($V_0 = 196.56 \us{V}$), $F_\x{ext} = 500 \us{MV}/\un{m}$ ($V_0 = 218.4 \us{V}$) and $F_\x{ext} = 550 \us{MV}/\un{m}$ ($V_0 = 240.24 \us{V}$). The anode is at a positive voltage $V_0$ and the cathode is grounded, resulting in the external field $F_\x{ext} = V_0 / h_\x{gap}$, with a local field $F_\x{loc} = \beta F_\x{ext}$, where $\beta$ is the (geometric) field enhancement. There is a serial RC circuit between the electrodes with resistance $R = 1\us{k}\Omega$ and capacitance $C = 1 \us{pF}$, but due to the short time scale and small size of the system, the voltage can be considered to be effectively constant. The work function for electron emission is $\phi = 4.5 \us{eV}$, with all particles (electrons, neutrals, ions) injected into the PIC system at superparticle weights of $w_\x{SP} = 1$. The time step for PIC is $\Delta t_\x{PIC} = 0.1 \us{fs}$ and the time step for the heat equation (and one whole FEMOCS step) is $\Delta t = 20 \us{fs}$. The bottom of the cathode is assumed to be at a temperature of $T_\x{amb} = 300 \us{K}$.

The MD system consists of all of the atoms in the nanoprotrusion, with the majority of the atoms thermostatted using the Berendsen thermostat (damping parameter $\tau = 1.5 \us{ps}$) based on the temperature distribution calculated from FEM. The bottom $100 \us{Å}$ layer is thermostatted in LAMMPS to 300~K to provide a smooth temperature transition to the static bulk. The bottom layer of atoms is fixed in place. The time step in LAMMPS is set dynamically every step based on atom velocities and forces, in the range $\Delta t_\x{MD} \in [10^{-6}, 10^{-2}] \us{ps}$. The length of the whole MD step is in total $\Delta t$.

\trifigw{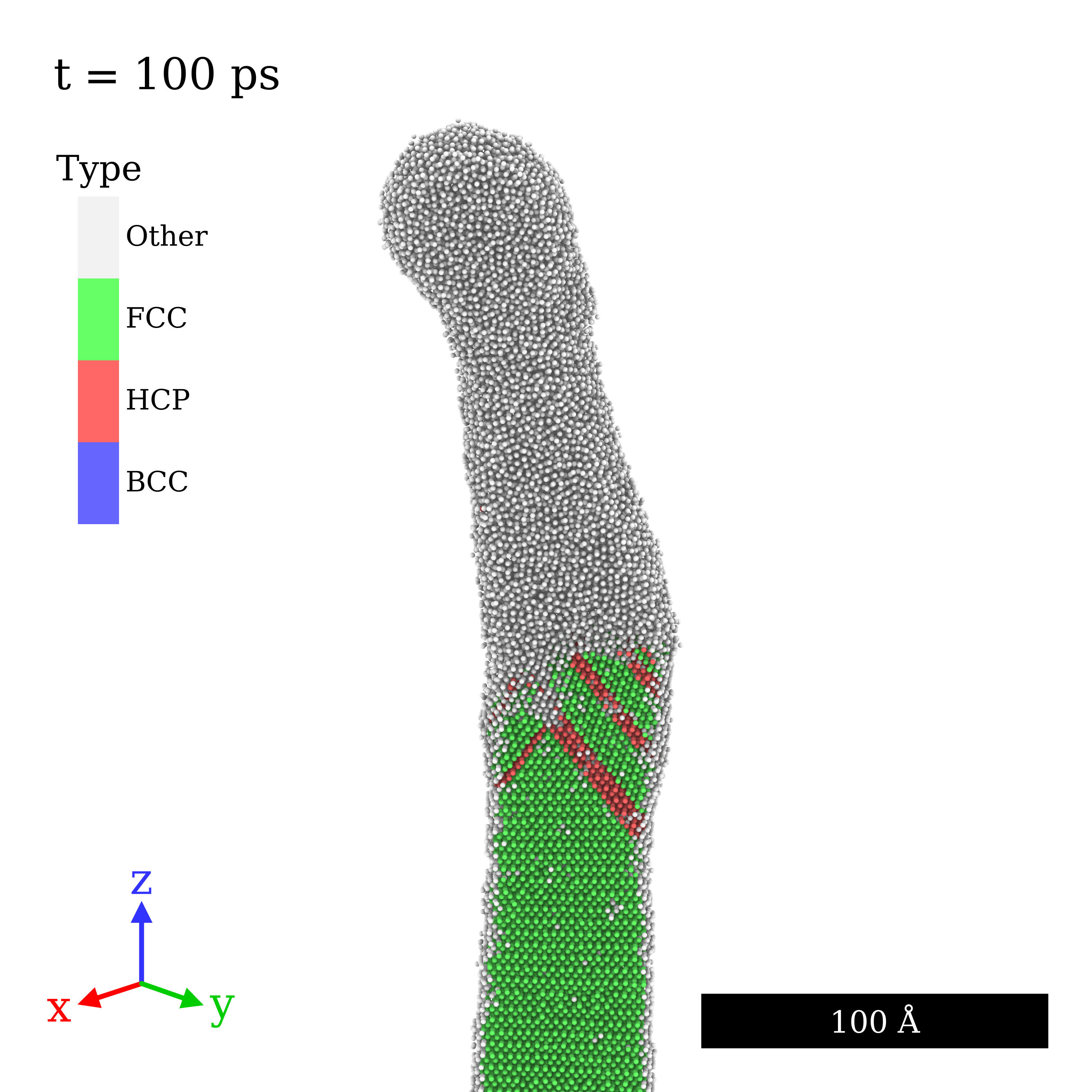}{Nanoprotrusion with $F_\x{ext} = 450 \us{MV/m}$, structure analysis shows crystalline and melted region.\label{fig:state450}}
{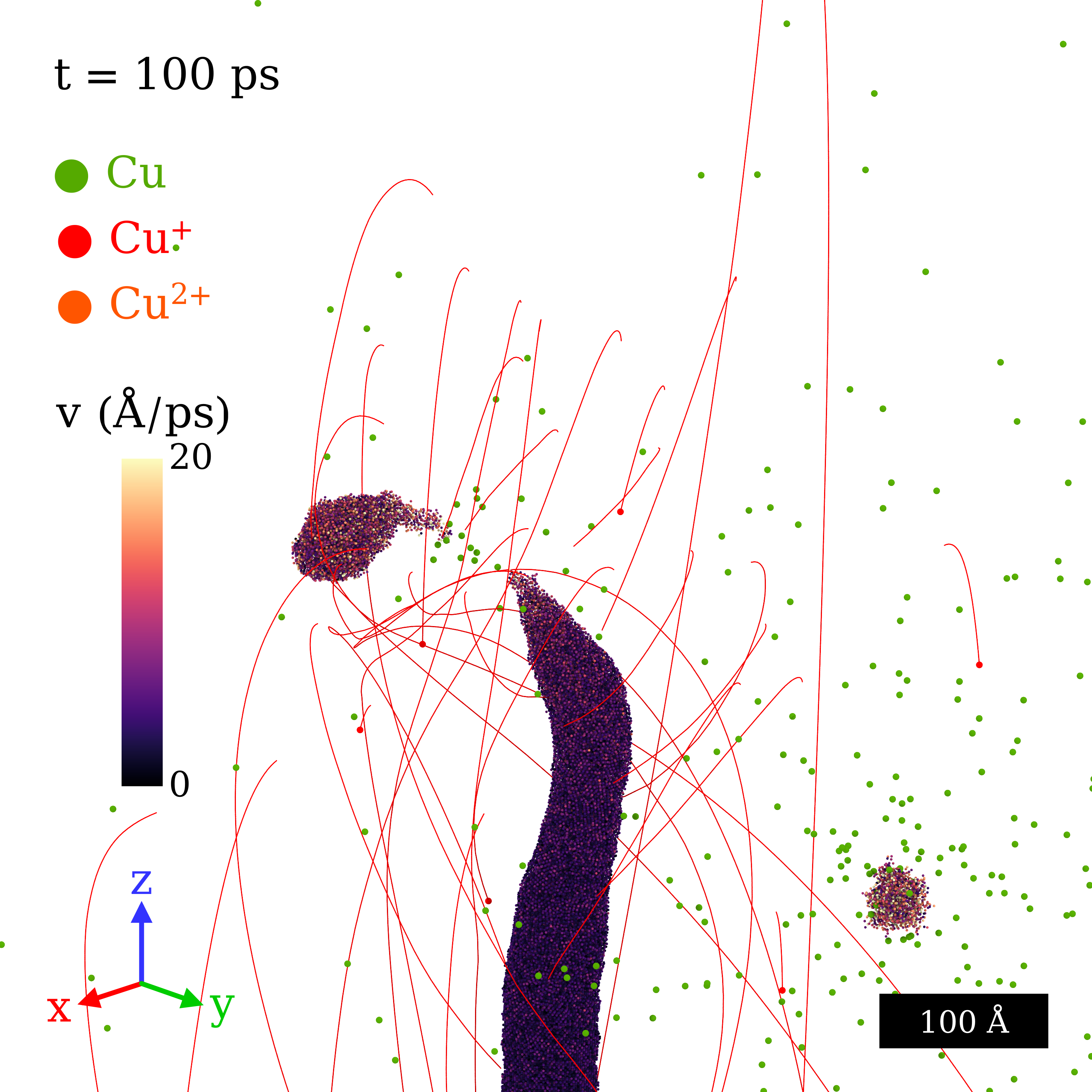}{Nanoprotrusion with $F_\x{ext} = 500 \us{MV/m}$, atoms colored by velocity, plasma particles indicated by legend and ion trajectories plotted as lines.\label{fig:state500}}
{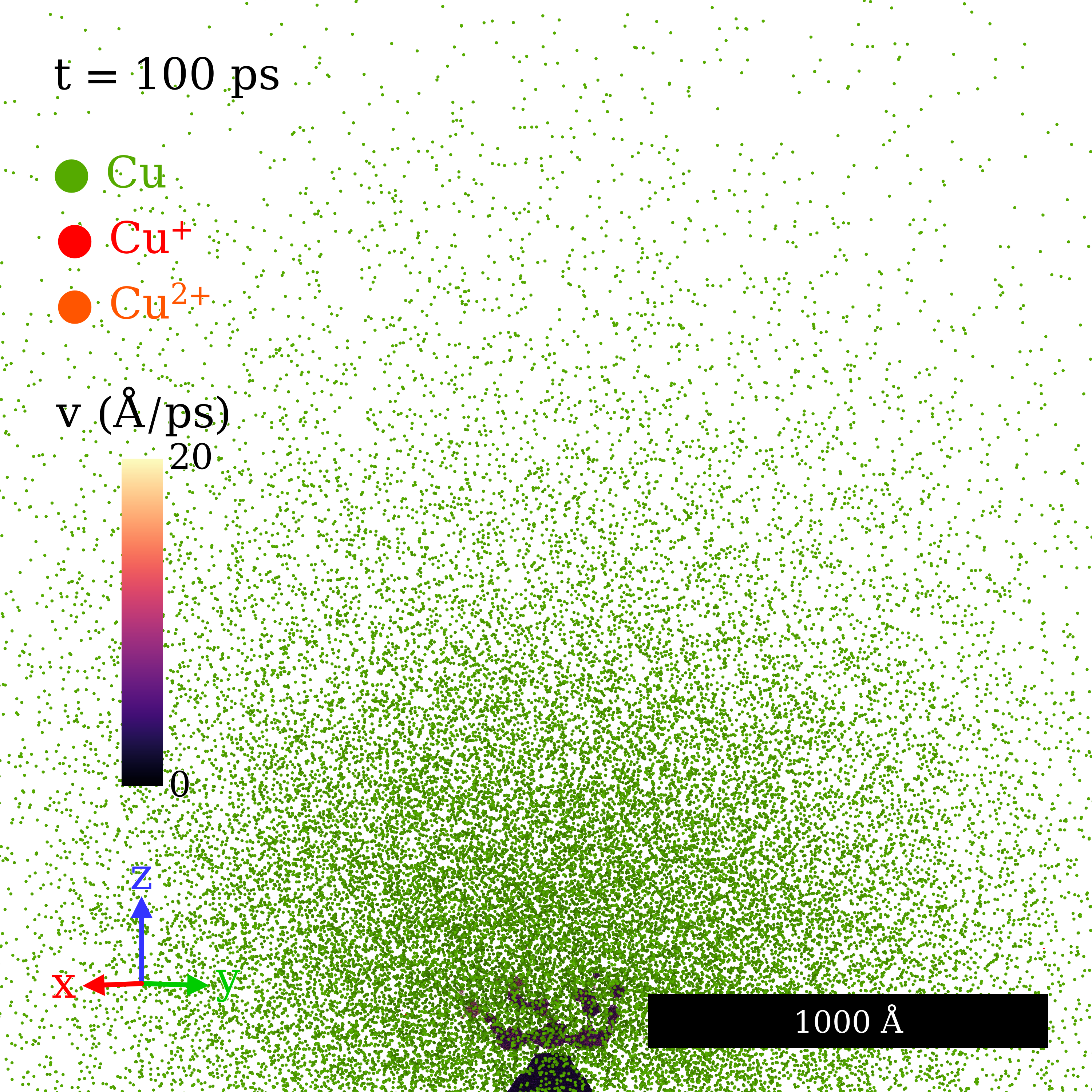}{Nanoprotrusion $F_\x{ext} = 550 \us{MV/m}$, remaining atoms and neutral vapor shown after runaway.\label{fig:state550}}
{Simulation state at $t = 100 \us{ps}$ for various field cases.}

In Figs.~\ref{fig:state450}--\ref{fig:state550}, the simulation state is shown for the various simulated field cases (visualized using OVITO~\cite{stu10}). For the lower field case with $F_\x{ext} = 450 \us{MV/m}$, the nanoprotrusion is shown to reach an apparent stady state, with an average maximum temperature of 827~K and maximum field of 12~GV/m over the last 50~ps. Structure analysis using polyhedral template matching (RMSD cutoff 0.1~Å) shows the crystal structure of the nanoprotrusion (Fig.~\ref{fig:state450}), with the top having an amorphous structure and the region below retaining its FCC structure. During the first 10~ps, the surface reached a maximum temperature of over 5000~K, causing the top to melt. The melted region collapsed due to surface tension overcoming the pulling electric forces, which resulted in a dramatic decrease in field enhancement, lowering the local maximum field from about 19~GV/m to 12~GV/m, which is on the edge of the expected threshold for arc initiation~\cite{des09}. The lower field resulted in decreased electron emission, which in combination with the small size of the nanoprotrusion did not generate enough Joule heating for thermal runaway to continue.

For the high field case with $F_\x{ext} = 550 \us{MV/m}$, thermal runaway occurred immediately due to a field that was high enough to sustain significant electron emission and Joule heating. The nanoprotrusion showed a high rate of evaporation and rapid reformation, with Cu NPs detaching off and disintegrating rapidly due to high temperature and additional electron heating. The whole structure disintegrated within 100~ps and was converted to neutral Cu vapor and Cu NPs (Fig.~\ref{fig:state550}).

The medium field case with $F_\x{ext} = 500 \us{MV/m}$ showed a slightly different thermal runaway mechanism compared to the high field case. Cooling of the surface due to the Nottingham effect resulted in relatively lower evaporation rates and lower emission currents slowed down heating and NP production. The main mechanism for neutral creation is evaporation from Cu NPs (Fig.~\ref{fig:state500}), which ionize and bombard the surface. Evaporation of the Cu NPs was observed to be a requirement for the thermal runaway to occur. An additional simulation with electron heat factor of $\alpha_\x{EH} = 0.1$ was performed, which showed no thermal runaway. Detachment of the NPs resulted in sufficient reduction in field enhancement that thermal runaway would not continue by itself within the simulated time. It was not until the NPs evaporated due to sufficient electron heat and ionized that an increase in current and temperature was seen, as shown in Fig.~\ref{fig:tcp}.

The mechanism for arc initiation by Cu NPs in the simulations is due to the formation of ions. The main factor influencing this thermal runaway pathway is the local electric field and electron emission. Once positively charged ions start forming above the surface, they cause an enhancement in the surface field, acting as a virtual electrode. This results in higher electron emission and Joule heating. Additionally, as ions are accelerated towards the surface and bombard it, they bring in heat and cause sputtering of more neutrals from the surface, reinforcing the process.

\dualfighs{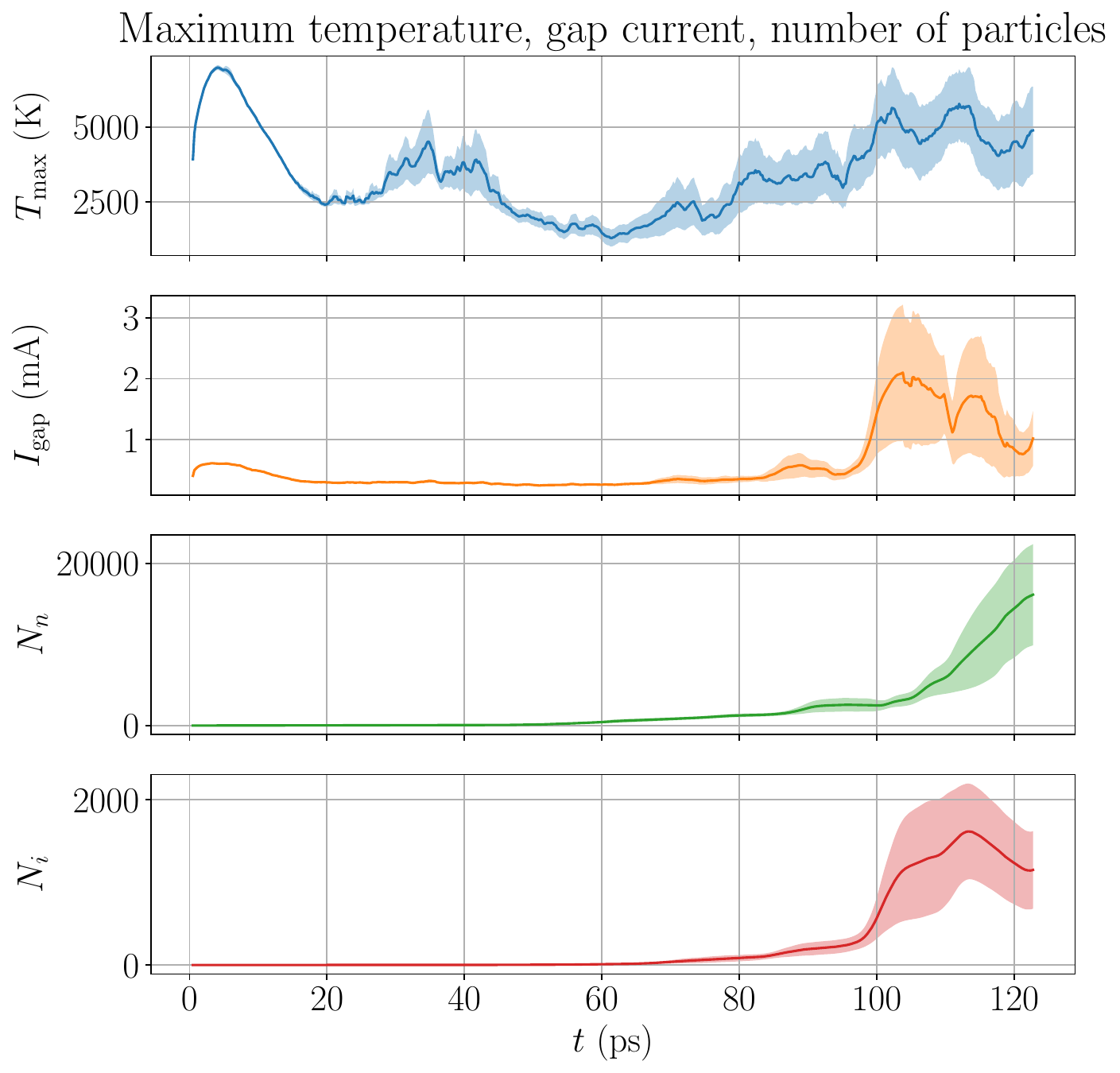}{Maximum surface temperature, gap current, number of neutrals and number of ions.\label{fig:tcp}}
{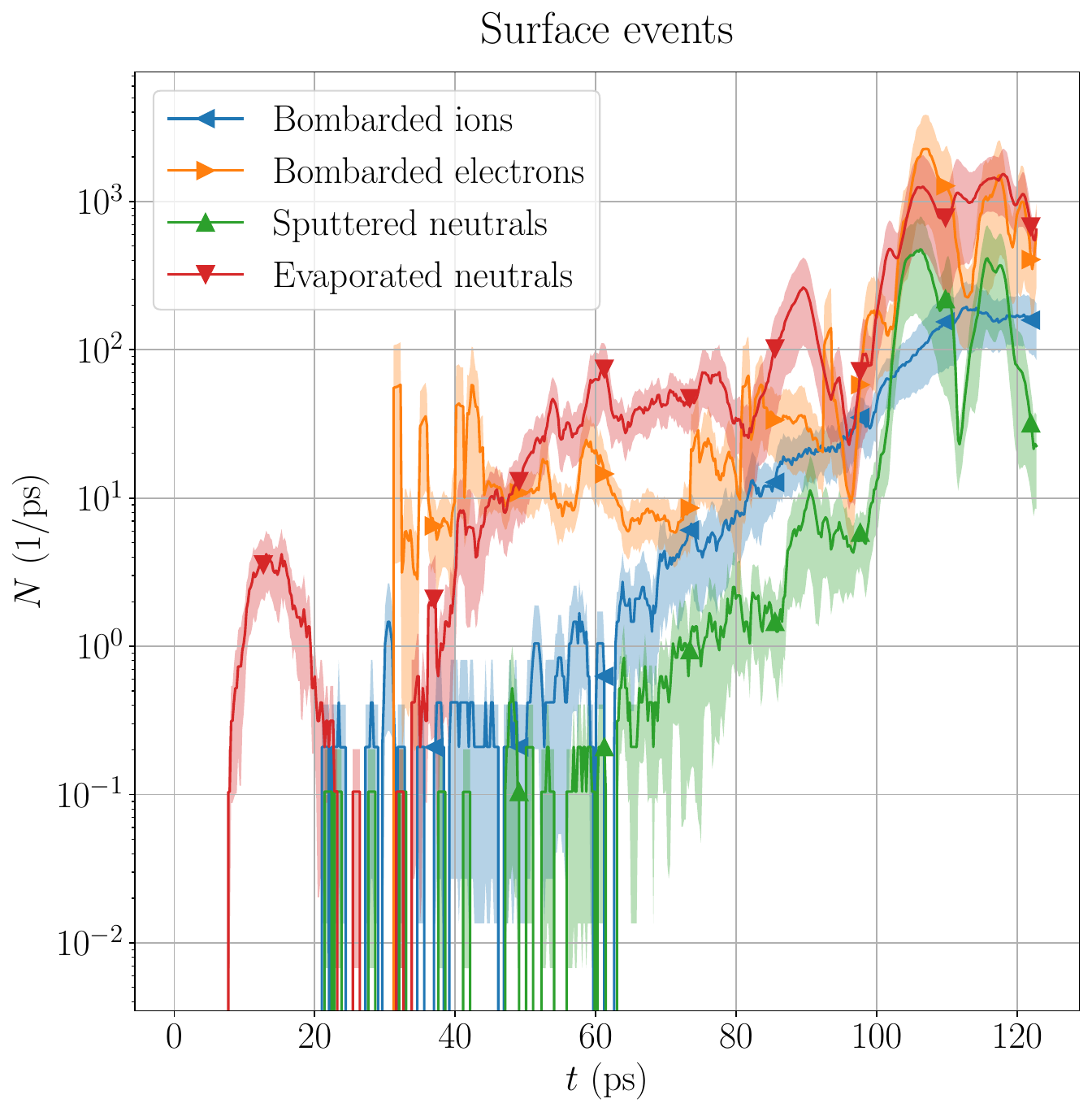}{Electron/ion bombardments of surface and sputtered/evaporated neutrals per unit of time.\label{fig:surf}}
{Averaged quantities of the simulated systems ($N=8$) as a function of time for the case with $F_\x{ext} = 500 \us{MV}/\un{m}$. The error bands indicate the standard error across runs.}{0.72}

In Fig.~\ref{fig:surf}, the cathode surface processes and their rates are shown. These include heating of Cu NPs via electron bombardment, ion bombardment of the cathode surface, sputtered atoms from the cathode surface and evaporated neutrals from the cathode surface and Cu NPs. The difference between evaporation and sputtering is detected based on the correlation of Cu atoms leaving the surface with ion bombardment. If a Cu atom leaves the surface and an ion bombardment event is detected within 1~ps and a cutoff distance of $10 v_n \Delta t + 5 a$, where $v_n$ is the neutral velocity and $a$ is the Cu lattice constant, it is classified as a sputtering event and evaporation otherwise. In Fig.~\ref{fig:surf}, it can be seen that electron bombardment and evaporation show a correlation, indicating that electron heating of Cu NPs plays a major role in creation of the initial neutral vapor. Additionally, there is a correlation between ion bombardment and sputtering, as would be expected. Each of the quantities show a similar trend as plasma buildup starts, suggesting that they are directly or indirectly connected to the formation of ions. Ion bombardment can influence neutral creation both directly via sputtering and bombardment heating (evaporation), but also indirectly due to the enhancement of the electric field and resulting electron emission.

\dualfighs{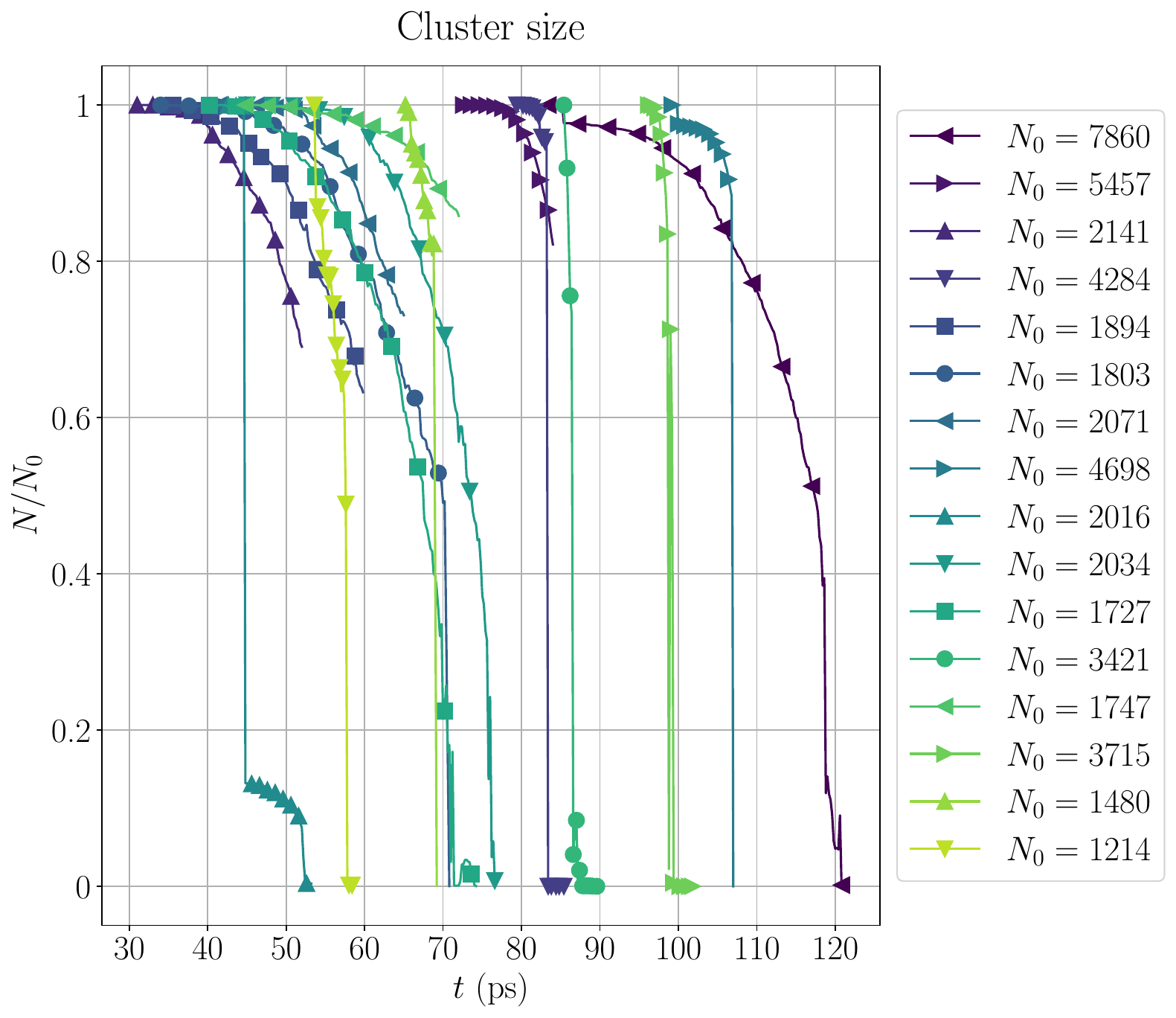}{Relative change in the number of atoms for each cluster.\label{fig:clsize}}
{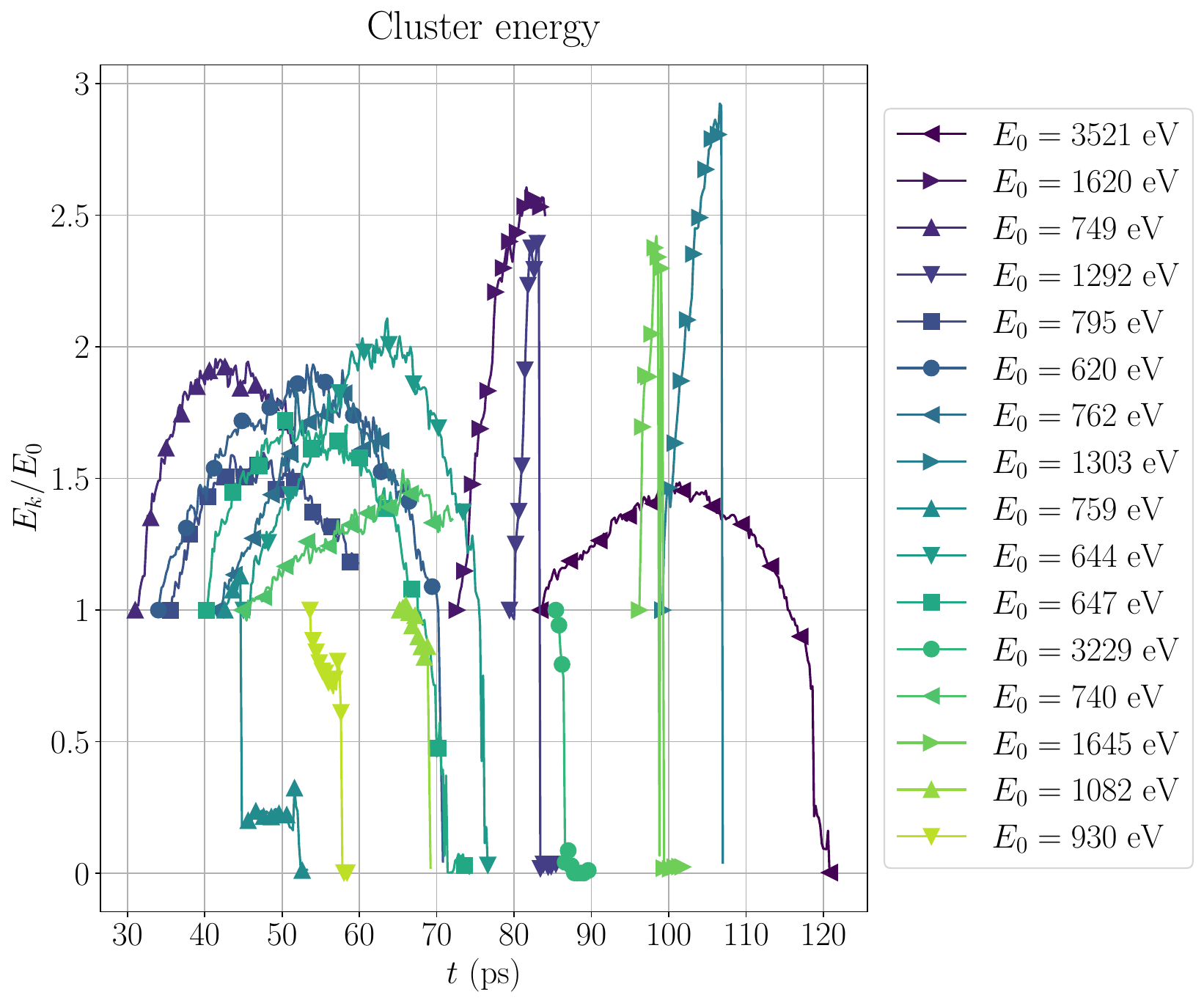}{Relative change in total energy for each cluster.\label{fig:clen}}
{Nanoparticles (NPs) in the simulated systems ($N=8$) over time for the case with $F_\x{ext} = 500 \us{MV}/\un{m}$. The labels indicate the initial number of atoms $N_0$ and initial total kinetic energy $E_0$ of each cluster.}{0.74}

In Figs.~\ref{fig:clsize} and \ref{fig:clen}, the Cu NPs are tracked across $N=8$ simulations, showing the relative changes in the number of atoms in each NP and in the total kinetic energy of each NP. Each cluster initially separates from the nanoprotrusion with size $N_0$ and total kinetic energy $E_0$, when they are removed from the FEM system and retained in the MD system. The clusters are no longer thermostatted, but they can gain energy via electron heating and lose energy via evaporation. Evaporation causes the NPs to shrink in size, which is shown in Fig.~\ref{fig:clsize}. When the NP initially separates, it receives a large amount of electron heating due to its proximity and size. As time goes on, the NP drifts further away, possibly out of the path of the electron beam. The size of the NP shrinks, decreasing its cross section and making it more unlikely to be hit. These factors can result in slowing down or stabilization of the NP size change. The time evolution of the NP energy in Fig.~\ref{fig:clen} shows the balance of electron heat compared to evaporative cooling. Initially, the energy is increasing when electron heating is more significant that evaporative cooling. Each of the largest NPs reaches a peak in energy, after which evaporation starts dominating. Between 70~ps and 80~ps, NPs are shown to break up very rapidly, correlating with an increase in emission current. At end of this process, a significant amount of neutral vapor and Cu NPs have been created.

We have demonstrated the fully coupled, dynamic thermal runaway of Cu nanoprotrusions and initial plasma precursor processes contributing to vacuum arcing and their mechanistic contributions to the phenomenon. As shown in previous work~\cite{kyr18,ves20}, thermal runaway of Cu nanoprotrusions can create Cu NPs such that a sufficient ratio of neutrals and electrons for plasma initiation could be created (assuming breakup of these NPs), motivated by plasma simulations~\cite{tim15}. The current work aims to unify these approaches into one, such that the processes and their couplings can be studied in a more complete manner. Our simulations show that the thermal runaway and plasma processes are intimately connected.

The simulations performed in this work suggest that there are two pathways to thermal runaway of nanoprotrusions: direct thermal runaway due to Joule heating and thermal runaway assisted by NP-created ions. The formation of a sheath of ions close to the cathode surface can significantly enhance electric fields, leading to increased emission and heating. Additional ion bombardment heating and sputtering would reinforce the process. Cu NPs can also act as a source for a large amount of Cu vapor between the electrodes, since they are likely to have large evaporation rates due to high temperatures resulting from thermal runaways, as well as additional electron heating.

The nanoprotrusions considered in this work are small enough that they likely do not have a sufficient number of atoms for the formation of a full plasma necessary to initiate a vacuum arc. There is the possibility that additional vapor from the flat cathode surface could be produced by sputtering of ions, which has been simulated previously at later stages of vacuum arcing~\cite{yan26}, though this would likely require a relatively large supply of ions. A possible hypothesis for the creation of the initial vapor is the appearance of a multitude of nanoprotrusions on the cathode surface. In such a scenario the cathode surface would see the growth, thermal runaway and extinguishment of many nanoprotrusions over a period of time. While each of these nanoprotrusions may not be sufficiently large to initiate the arc on their own, they contribute to the total supply of neutral vapor. Once a sufficient amount of vapor has been accumulated, one such thermal runaway event may act as the trigger for an avalanche of ionizations, creating the required number of ions for a feedback process to occur. Such ionization avalanches have been observed in previous simulations~\cite{koi24}, where the production of secondary electrons results in additional impact ionizations.

We have described the modeling of the vacuum arc initiation process using multiphysics simulations that bridge the gap between atomistic cathode surface processes and plasma initiation. An interesting direction for future work is the application of this model to different materials and geometries, allowing detailed study of the properties that can factor into breakdown thresholds.



RK, AK, TT and VZ are supported by the European Union's Horizon 2020 research and innovation program, under Grant Agreement No. 856705 (ERA Chair ``MATTER'') and by the Estonian Research Council's grants PRG2675, TARISTU24-TK10 and TEM-TA23. TT is supported by the Estonian Research Council Grant No. SJD61. RK, AK and FD have been supported by CERN CLIC K-contract (No. 47207461). RK is supported by the doctoral program MATRENA of the University of Helsinki. The authors wish to thank the Finnish Computing Competence Infrastructure (FCCI) for supporting this project with computational and data storage resources.






\bibliography{vacuum_arc_plasma_surface}

\appendix

\section{Simulation model description}\label{sec:model}

\fig{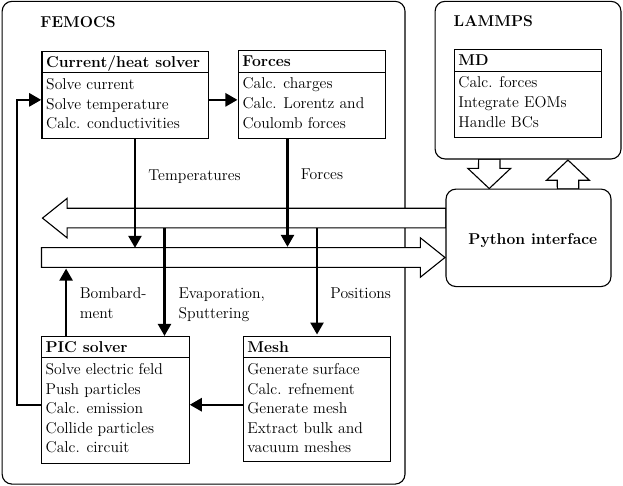}{Flowchart showing the coupling between the FEM+PIC simulation (FEMOCS) and the MD simulation (LAMMPS).\label{fig:model}}{0.9}

\noindent The coupled simulation is a combination of the FEM+MD simulation~\cite{ves20} and FEM+PIC simulation~\cite{koi24}. A flowchart showing how parts of the simulation model are connected is shown in Fig.~\ref{fig:model}. The course of the coupled FEM+PIC+MD simulation can be summarized as follows:
\begin{enumerate}
\item Create FEM mesh based on atom positions
\item Transfer particles from MD to PIC
  \begin{enumerate}
  \item Evaporated atoms
    \item Sputtered atoms
  \end{enumerate}
\item Run FEM+PIC step for $\Delta t$ (FEMOCS)
  \begin{enumerate}
  \item Electric field, electron emission (GETELEC)
  \item PIC step (push particles and perform collisions)
  \item Current and temperature distribution
  \item Surface charges (forces)
  \end{enumerate}
\item Transfer FEM+PIC quantities to MD
  \begin{enumerate}
  \item Surface charges $\rightarrow$ atom forces
  \item Temperatures $\rightarrow$ atom velocities
  \item Electron heat (NPs)
  \item Bombarding ions
  \end{enumerate}
\item Run MD step for $\Delta t$ (LAMMPS)
\end{enumerate}

Coupling of the FEMOCS and LAMMPS simulations is performed using a Python interface, which allows communication of atom quantities between the two. The FEMOCS and LAMMPS simulation steps are run in a staggered fashion, alternating between advancing the FEM+PIC system by $\Delta t$ and advancing the MD system by $\Delta t$.

Meshes for the vacuum domain and bulk domain are generated based on atom positions by detecting the atomic surface and separating the mesh into two pieces using a ranking procedure. Surface detection is based on calculation of the number of nearest neighbors and a coordination number cutoff (5.75~Å). Mesh quality is defined for different regions, with the finest mesh density at the top of the nanoprotrusion and an outwardly coarsening mesh on the flat bulk surface. Mesh quality is controlled using maximum volume and element aspect ratio parameters, with automatic mesh refinement in the vacuum domain based on the distribution of plasma. Debye length (for electrons + ions) is calculated for each vacuum element, which is used to calculate the required size~\cite{tsk07} $d_\x{max} = 3.4 \lambda_D \alpha_\x{D}$ with tolerance $\alpha_\x{D} = 0.5$. Additionally, we determine if the mesh needs to be recalculated based on the root mean squared distance of the atoms exceeding the tolerance limit $\delta_\x{RMSD} = 0.5 \us{Å}$. Cluster analysis is performed with a cutoff of 4.1~Å to detect atoms that have left the cathode surface and excluded from the mesh.

There are three atom types in the MD system: normal Cu atoms (type 1), bombarding ions (type 2) and ghost atoms (type 3). Type 1 and type 2 atoms interact with all atoms of type 1 or 2 via an EAM potential for copper by Mishin et al.~\cite{mis01}. Type 3 atoms don't interact with any atoms of types 1, 2 or 3. Only type 1 atoms are included in the FEM mesh building process, while type 2 and type 3 atoms are excluded from it, though only type 1 atoms part of the main cluster connected to the cathode bulk are included in the mesh. Likewise, only type 1 atoms are thermostatted by the Berendsen thermostat. Type 2 atoms are used for particles that are impacting the surface, so that they are not included in the mesh or thermostatted, which could alter their energy. Type 3 atoms are used for particles that leave the surface, such as evaporated atoms or sputtered atoms.

When a single atom or a cluster with a small size (less than 10) separates from the rest, it is considered as evaporated, after which it is removed from the MD system by changing its type to a ghost atom and added to the PIC system as a superparticle with weight 1. Ions impacting the cathode surface in the PIC system are added back to the MD system as type 2 atoms such that their position is offset backwards along their velocity vector by 3 lattice constants. This ensures that the bombarding atoms are not added inside the surface or overlapping with surface atoms. When the bombarding atom is detected as part of the same cluster as the main one, its type is set to 1. Atoms that may sputter from surface as a result of such impacts are transferred from the MD system to the PIC system similarly as evaporated ones.

When large clusters of atoms, i.e. nanoparticles (NPs) detach from the nanoprotrusion, they are retained in the MD system, but excluded from FEM. The NPs separate from the nanoprotrusion in the molten state, so they stay at a high temperature and keep evaporating atoms. Such NPs can lose energy by evaporative cooling due to evaporation of atoms or by radiation, which is relatively inefficient. There is also the possibility to gain energy due to interaction with the emitted electron beam. When electrons impact the Cu NP, they can lose some of they energy due to electron-phonon scattering. This interaction is included in the model using an electron energy loss fraction parameter $\alpha_\x{EH}$. When an electron with energy $E_e$ hits the NP, $\alpha_\x{EH} E_e$ of its energy is deposited to the whole NP in the form of heat. This heat is added to each atom in the NP by applying an impulse in a random spatial direction such that the energy would increase by $\alpha_\x{EH} E_e / N_c$ if starting from rest, where $N_c$ is the number of atoms in the cluster. The probability for an electron to pass through the NP can be estimated from the inelastic mean free path, which is $\ell \approx 7 \us{Å}$ for 20~eV electrons (average energy from simulation) in Cu~\cite{dev19}. There is also the possibility for the electron to be reflected away from the NP, which is estimated to be $R \approx 0.15$~\cite{mcr76}. The energy loss parameter is estimated to be $\alpha_\x{EH} = (1 - R) \exp(-d/\ell) (1 - \phi / E_e) = 0.85 \times 0.76 \times (1 - 4.5 / 20) \approx 0.5$ for a Cu NP with a diameter of $d=1\us{nm}$ (about half of the nanoprotrusion), such that each electron that would be lost leaves the NP with an energy equal to the work function $\phi$. Under this approximation, each electron gives on average 50\% of its energy to the NP. Since the effect of temperature on the inelastic mean free path, as well as partial energy transfer of transmitted electrons is ignored, this is likely a conservative estimate of the deposited energy.

\end{document}